\documentclass[aps,prl,twocolumn,superscriptaddress,nofootinbib,
nobalancelastpage,showpacs,nobibnotes]{revtex4}
\usepackage{epsfig}

\begin{document}
\title
{ The multi-angle instability in dense neutrino systems 
}

\author{R. F. Sawyer}
\affiliation{Department of Physics, University of California at
Santa Barbara, Santa Barbara, California 93106}

\begin{abstract}

We calculate rates of flavor exchange within clouds of neutrinos
interacting with each other through the standard model 
coupling, assuming a conventional mass matrix.  For cases in which 
there is an angular dependence
in the relation among intensity, flavor and spectrum, we find instabilities
in the evolution equations and greatly speeded-up flavor exchange.
The instabilities are categorized by examining linear perturbations
to simple solutions, and their effects are exhibited in complete
numerical solutions to the system.
The application is to the region just under the neutrino surfaces 
in the supernova core.


\end{abstract}
\maketitle

\section{1. Introduction}
There are least two astrophysical contexts in which the standard model interaction of 
neutrinos with neutrinos (mediated by Z exchange) could play a important role:

1). In the supernova process, either in the region just under the neutrinosurface (i.e. the 
point of last scattering, on the average), or in the regions just outside of this region.
The energy distributions of the various species of neutrino in this region are important
to the dynamics of the explosion, to R process nucleosynthesis downstream, and
to the observed neutrino pulse on earth, should it ever be observed again. These energy
distributions may be affected greatly by coherent neutrino-neutrino interactions, even though
the effects of ordinary incoherent $\nu-\nu$ scattering are small. There has been much recent 
work on the combined effects of $\nu$-$\nu$
interactions and oscillations in the region outside the neutrino surfaces
for neutrinos of different flavors\cite{ful1}-\cite{ful5}.

2). In models of the early universe before nucleosynthesis (but after leptogenesis, etc.
and well before $\nu$ decoupling) in which there are substantial neutrino chemical potentials,
that is to say, neutrino anti-neutrino imbalance at much greater level than the electron-
positron imbalance. If flavors are out of equilibrium as well, then $\nu-\nu$ interactions become important.
When sterile neutrinos are added, these scenarios become arbitrarily
complex. They can easily change the parameters that determine the He 
abundance, so that a part of the effort expended on their behalf has been
to avoid upsetting the good fits of the standard theory. The models can be arranged 
to produce models of sterile neutrino dark matter in the mass region at least several KeV,
coupled with a mixing parameter small enough not to conflict with cosmic x-ray backgrounds
\cite{gf}-\cite{gf3}.

The present work operates in the context of the supernova problem,
although there could well be applications to the area b), above,
as well. 
Here we consider some effects below the neutrino surface, or, more accurately,
in the transitional region at about the depth of the average last scattering for
$\nu_{\{\mu, \tau\}}$ and $\bar \nu_{\{\mu, \tau\}}$, and below the depth
of the average last scattering for
$\nu_{e}$ and $\bar \nu_{e}$.
This follows up on previous work \cite{rfs1}, \cite{rfs2} in the same region, demonstrating the possibility
of flavor instabilities causing rapid mixing of the angular distributions
and energy spectra. In these instabilities, which depend only on forward
$\nu$ processes, flavor is traded between momentum states with no change
in the momentum states. The instabilities arise when neutrino momentum
distributions are anisotropic. The examples worked out in refs. \cite{raf3},\cite{raf5}  dealing with a region outside of the neutrino-surfaces have at their core the type of instability
as that described in \cite{rfs1}, \cite{rfs2}, which we characterize as the ``multi-angle instability."

The present work has three purposes:

a) To provide a better basis for understanding these instabilities, described
in the above references simply in terms of the plots produced in simulations.
In the present paper we carry out standard linearized stability analysis of the underlying
nonlinear systems.

b) To do a more realistic analysis than was done in ref. \cite{rfs1} of flavor-scrambling effects in the region just
under the neutrino surfaces. For example,
in the current work we show results for a case with 14 bins of $\nu$ momentum space,
as compared to 2 bins in ref. \cite{rfs1}

c) To point out a second category of possible instability, potentially more potent than
the one underlying the work mentioned above. 
A feature of the results is that small fluctuations of initial distributions away from symmetrical
cases can have a catalyzing effect, inducing complete flavor transfers between
up-moving and down-moving states.

Our effects evolve over small distances in homogeneous matter, in contrast
to MSW resonance flavor trading, which depends on passage through more extended regions 
with variable electron density. They also act to mix up neutrino and anti-neutrino
flavors simultaneously, in contrast to resonant phenomena, which act on one or the other. 

The medium
in the region of the neutrinosurfaces has density of a few times $10^{11}\,{\rm  g\,\rm c}^{-3}$ at a
temperature $T=5-7 \,{\rm MeV}$.
We shall consider neutrino mixing of two species, $\nu_e$ and (following the notation of
other authors) $\nu_x$, where $\nu_x$ is some mixture of  $\nu_\mu$ $\nu_\tau$.
There are a number of time scales that we can define in this region:

1. A ``very fast" time scale $(G_F n_\nu)^{-1}$, where $n_\nu$
is the neutrino number density $n_e \sim T^3$. In our domain this time  
is of order $10^{-3}\rm cm.$ in units in which $c=1$. 

2. The vacuum oscillation time, of order $10^7$ cm. in our region, for
$\delta m^2=.8 \times 10^{-4} {\rm (eV)}^2$.
All of the rates of change for interesting quantities that we calculate in 
the present paper will be much faster than this rate.

3. A ``fast" rate that is the geometric mean of the above two rates. The fact that 
properties of $\nu$ systems can change at this rate was noted by Kostelecky and Samuel
\cite{ks} \cite{sam} in their work on self-maintained coherent oscillations in dense isotropic neutrino
gases. More recent exploration of this topic can be found in
refs. \cite{raf4} and \cite{raf2}. This basic rate recurs in the multi-angle flavor instabilities
found in refs. \cite{raf3},\cite{raf5},\cite{rfs1}. 

4. The rate for scattering from neutrons and protons in the medium $\sim G_F^2 T^3$
about ten times as fast as the vacuum oscillation rate in the domain of interest.

The equations for flavor density matrices needed to derive our principal 
results are standard by now \cite{raf1}. But in order to explain some caveats in a
later section and to set the groundwork for future work we shall briefly rederive the standard equations. 

We consider a system of neutrinos interacting with each other 
through the Z mediated neutral current couplings, and with electrons and positrons
in the medium, through the W and Z mediated couplings.  Interactions of $\nu$'s with nucleons
in the medium are irrelevant, as far as the main developments are concerned, since they
are independent of $\nu$ flavor. We also include ordinary neutrino oscillations.
The effects of ordinary incoherent scattering by nucleons and electrons in the
medium can be neglected over the time scales that we consider. 

We define $a_e({\bf p})$ and $a_x({\bf p})$, as the respective annihilation
operators for a $\nu_e$ and $\nu_x$ of momentum $\bf p$, with corresponding definitions
for the operators that annihilate $\bar \nu$'s. 
We introduce the density operators, 

\begin{eqnarray}
\rho_{i,j}({\bf p})=a_i({\bf p})^\dagger \, a_j({\bf p})
\nonumber\\
\bar \rho_{i,j}({\bf p})= \bar a_j({\bf p})^\dagger \,\bar a_i({\bf p}) \, ,
\end{eqnarray}
where $i$ and $j$ take the values $e$ or $x$. Note the transposition of indices
in the definition of the antiparticle density operator; the leads to neater
formulae below.  

The reason that $\nu$-$\nu$ interactions can change flavor-spectrum correlations
over a very short time scale is that in a pure forward encounter between two neutrinos
the neutrinos can swap flavors, thus engendering coherent (because of forward) effects
that are more than phases in their wave-functions (because of flavor dependence). We start therefore by
isolating from the Z exchange coupling all of the terms with no momentum exchange, that is, 
the terms that connect
a pair of $\nu$'s with momenta $\bf p,q$ to a pair with momentum $\bf p,q$; only the flavors being
changed. We shall refer to this as the ``forward" neutrino-neutrino interaction, $H_{\nu,\nu}$ ,
\begin{eqnarray}
&H_{\nu \nu}(\rho)= {\sqrt{2}G_F\over V }\sum_{{\bf p, q} }~\sum _{\{i,j\}=e,x}
 [1-\cos (\theta_{\bf p,q})]
\nonumber\\
&\times \Bigr[ \Bigr(\rho_{i,j}({\bf p})-\bar \rho_{i,j}({\bf p})\Bigr )
\Bigr(\rho_{j,i}({\bf q})-\bar \rho_{j,i}({\bf q})\Bigr )~~~~~~~~~~
\nonumber\\
&~~~~~+\Bigr(\rho_{i,i}({\bf p})-\bar  \rho_{i,i}({\bf p})\Bigr )
\Bigr(\rho_{j,j}({\bf q})-\bar \rho_{j,j}({\bf q})\Bigr )\Bigr ]\, ,
~~~~~~~~~~~~~~~~
\label{ham}
\end{eqnarray}
where $V$ is the volume.

The oscillation (or mass-matrix) term $H_{\rm osc}$
is taken to be of the form,
\begin{eqnarray}
&~~~~H_{\rm osc}(\rho)=
\nonumber\\
&\sum_{\bf p}  \Bigr [ |{2\bf p}|^{-1} \eta [\rho_{e,x}({\bf p})+\rho_{x,e}({\bf p})
+\bar\rho_{e,x}({\bf p})+\bar \rho_{x,e}({\bf p})]
\nonumber\\
&+|{2\bf p}|^{-1}\xi [  \rho_{e,e}({\bf p})-\rho_{x,x}({\bf p})+ \bar \rho_{e,e}({\bf p})-\bar \rho_{x,x}({\bf p})] 
\nonumber\\
&+G_F n_e [  \rho_{e,e}({\bf p})-\rho_{x,x}({\bf p})- \bar \rho_{e,e}({\bf p})+\bar \rho_{x,x}({\bf p})] 
   \Bigr ]\, ,
\label{osc}
\end{eqnarray}
where the parameters $\xi$, $\eta$ determine the standard vacuum neutrino mass parameters
according to $\delta m^2=(\xi^2+\eta^2)$, $ \tan (\theta)=\eta /\xi$, and the third line in (\ref{osc}) comes
from the standard model neutrino electron interaction.

In the main part of what follows, we discuss the time evolution of a neutrino
system under the influence of the Hamiltonian $H_1$, where,
\begin{eqnarray}
H_1(\rho)=H_{\rm \nu \nu}(\rho)+H_{\rm osc}(\rho)
\label{totalham}\, .
\end{eqnarray}
Since $H_1(\rho)$ in (\ref{totalham}) leaves the momenta of the individual neutrinos
unchanged, the kinetic energy term (in the absence of neutrino mass)
is irrelevent to evolution under the influence of $H_1(\rho)$.

The commutation rules of the density operators are
\begin{eqnarray}
[\rho_{i,j}({\bf p}),\rho_{k,l}({\bf p'})]=[\delta_{i,l}\rho_{k,j}({\bf p})-\delta_{j,k}\rho_{i,l}({\bf p})]\delta_{\bf p,p'}\,,
\nonumber\\
\,
\nonumber\\
\,[\bar \rho_{i,j}({\bf p}),\bar \rho_{k,l}({\bf p}')]=[-\delta_{i,l}\bar\rho_{k,j}({\bf p})+\delta_{j,k}\bar\rho_{i,l}({\bf p})]\delta_{\bf p,p'}\,.
\label{com}
\end{eqnarray}
With this apparatus the usual equations for the density functions are of the Heisenberg form,

\begin{eqnarray}
i {d \over dt} \rho_{i,j}({\bf p})=[\rho_{i,j}({\bf p}),H]\, .
\label{heisenberg}
\end{eqnarray}
We obtain,

\begin{eqnarray}
&{d \over dt}\rho_{i,j} ({\bf p})={-\sqrt{2} G_F \over V} \sum_{\bf q}\sum_k 
\Bigr[\rho_{i,k}({\bf p})
[\rho_{k,j}({\bf q})-\bar \rho_{k,j}({\bf q})]
\nonumber\\
&- \rho_{j,k}({\bf p})[ \rho_{i,k}({\bf q})- \bar \rho_{i,k}({\bf q})] \Bigr ][1-\cos (\theta_{\bf p,q})]
\nonumber\\
&+|{\bf p}|^{-1}[\Lambda,\rho ({\bf p})]_{i,j} \,,
\label{eom0}
\end{eqnarray}
and, 
\begin{eqnarray}
&{d \over dt}\bar\rho_{i,j} ({\bf p})={-\sqrt{2} G_F \over V} \sum_{\bf q}\sum_k 
\Bigr[\bar \rho_{i,k}({\bf p})
[\rho_{k,j}({\bf q})-\bar \rho_{k,j}({\bf q})]
\nonumber\\
&-\bar \rho_{j,k}({\bf p})[ \rho_{i,k}({\bf q})- \bar \rho_{i,k}({\bf q})] \Bigr ][1-\cos (\theta_{\bf p,q})]
\nonumber\\
&+|{\bf p}|^{-1}[\Lambda,\bar \rho ({\bf p})]_{i,j} \, .  
\label{eom0b}
\end{eqnarray}

For brevity we expressed the part coming from the commutator with $H_{\rm osc}$ in
(\ref{osc}) terms of a conventional generator $\Lambda (\lambda,\xi)$, a 2$\times$2 matrix in the
flavor space.
\section{2. Mean field equations}
In this paper we concentrate on a ``mean-field" limit, in which the operators (in the
occupation number space), $\rho_{i,j}$, are replaced by their expectation values 
in the medium, $<\rho_{i,j}>$, {\it after} the commutators (in the occupation number
space) have been performed to give the evolution equations (\ref{eom0}),(\ref{eom0b}).

This mean field assumption can be stated as
\begin{eqnarray}
\langle \rho_{i,j}(p) \rho_{k,l}(p')\rangle=\langle \rho_{i,j}(p) \rangle \langle \rho_{k,l}(p') \rangle \,.
\label{mf}
\end{eqnarray}
In this case we can suppress the indices $<\rho_{i,j}>$ and write the evolution
equations (\ref{eom0}) and (\ref{eom0b}) as matrix equations in the flavor indices,
generated in canonical fashion by effective mode by mode Hamiltonians for the matrices
$\rho(p)$,
\begin{eqnarray}
&H_{eff}({\bf p})={\sqrt{2} G_F \over V}\sum _{\bf q}[ \rho ({\bf q})- \bar \rho ({\bf q})][1-\cos (\theta_{\bf p,q})]
\nonumber\\
&+|{\bf p}|^{-1}\Lambda \, .
\label{mfham}
\end{eqnarray} 

We emphasize that the formulation based on (\ref{mfham}) depends totally on the
mean field assumption. Although it probably gives the correct answers in the problems
that we address, we shall return to discuss possible corrections to the mean-field
approximation. This discussion
must be based on (\ref{ham}) rather than on (\ref{mfham}).

For computational purposes we introduce collective variables for neutrinos by subdividing 
the momentum space into some small number $N_{\rm B}$ of regions $\{\alpha\}$ which we designate as
beams. For convenience we will do the
subdivision so that there are the same number of neutrinos + antineutrinos, $N_\nu$, in each beam $\alpha$,
\begin{equation}
N_\nu=\sum_i \sum_{{\bf p} \subset \{\alpha \}} [\rho_{i,i}({\bf p})+\bar \rho_{i,i}({\bf p})]
=n_\nu V/N_{\rm B}\, .
\label{Ndef}
\end{equation}
Then we define density matrices for the individual beams, 
\begin{eqnarray}
\rho^{(\alpha)}_{i,j} =N_\nu^{-1}\sum_{{\bf p} \subset \{\alpha \}}\rho_{i,j}({\bf p})\,,
\label{rhobeam}
\end{eqnarray}
and
\begin{eqnarray}
\bar \rho^{( \alpha)}_{i,j} =N_\nu^{-1}\sum_{{\bf p} \subset \{\alpha \}}\bar \rho_{i,j}({\bf p})\,,
\label{rhobeam}
\end{eqnarray}
so that in each beam, $\alpha$, we have,
\begin{eqnarray}
\sum_{j}[\rho^{(\alpha)}_{j,j}+\bar\rho^{(\alpha)}_{j,j}]=N_B^{-1}\, .
\end{eqnarray}

For calculations we need only two operators in each beam for particles
and two for antiparticles. We define,
\begin{eqnarray}
r_\alpha=\rho_{e,x}^{(\alpha)} ~~,  \bar r_\alpha=\bar \rho_{e,x}^{(\alpha)},
\nonumber\\
z_\alpha=\rho_{e,e}^{(\alpha)}-\rho_{x,x}^{(\alpha)}~~ ,~~\bar z_\alpha=\bar \rho_{e,e}^{(\alpha)}-\bar \rho_{x,x}^{(\alpha)}\, .
\label{redef}
\end{eqnarray}
In the two flavor case it has become conventional to write the evolution
equations for matrices that are hermitean 3-vectors in the internal flavor space. But it is computationally more efficient to
use, for each value of $\alpha$, non-hermitian $r_\alpha$'s, introduced above, rather than the 
two hermitian components.

The effective Hamiltonian is now,
\begin{eqnarray}
&H={\sqrt{2}G_F N_\nu n_\nu \over N_{B} } \sum_{\alpha,\beta} \Bigr [ 2 (r_\alpha-\bar r_\alpha)^\dagger (r_\beta-\bar r_\beta) 
\nonumber\\
&+(z_\alpha-\bar z_\alpha) (z_\beta-\bar z_\beta)\Bigr ]
\times[1-\cos\theta_{\alpha,\beta}]
\nonumber\\
 &+N_\nu \sum_\alpha \Bigr[ {|\bf p_\alpha}|^{-1}\lambda ( r_\alpha +r_\alpha^\dagger +\bar r_\alpha+\bar r_\alpha^\dagger)+
{|\bf p_\alpha}|^{-1} \xi (z_\alpha  +\bar z_\alpha)
\nonumber\\
&+G_F n_\nu (z_\alpha -\bar z_\alpha)\Bigr ] \, .
\label{ham2}
\end{eqnarray}
We have here dropped  in (\ref{ham}) the terms that involve only $\sum _i \rho_{i,i}(p)$ and $\sum _i \bar \rho_{i,i}(p)$,
since these quantities are conserved under the present dynamics \footnote{These terms
would enter the dynamics if, for example, we were including mixing with sterile neutrinos.} 

From (\ref{com} )and (\ref{redef}) the commutation rules of our dimensionless variables are
\begin{eqnarray}
[r_\alpha, r_\beta^\dagger]=N_\nu^{-1} z_\alpha \,\delta_{\alpha, \beta}~~,~~[r_\alpha, z_\beta]=2
N_\nu^{-1}r_\alpha \, \delta_{\alpha,\beta} 
\nonumber\\
\,[\bar r_\alpha, \bar r_\beta^\dagger]=-N_\nu^{-1}\bar z_\alpha\,\delta_{\alpha, \beta} ~~,~~[\bar r_\alpha, \bar z_\beta]=-2 N_\nu^{-1} r_\alpha \, \delta_{\alpha,\beta}\, .
\nonumber\\
\,
\label{coms}
\end{eqnarray}
The equations for time evolution are given for the flavor changing
operators by,
\begin{eqnarray}
&i {d \over dt} r_\alpha ={ \sqrt{2} G_F n_\nu\over N_{\rm reg}}\sum_\beta\Bigr [z_\alpha (r_\beta-\bar r_\beta )-r_\alpha (z_\beta-
\bar z_\beta)\Bigr] 
\nonumber\\
&\times[1-\cos \theta_{\alpha ,\beta }] 
+(\eta z_\alpha  - \xi r_\alpha ) /|{\bf p_\alpha}|+G_F n_e r_\alpha\,,
\nonumber\\
\,
\nonumber\\
&i {d \over dt} \bar r_\alpha =-{ \sqrt{2} G_F n_\nu\over N_{\rm reg}}\sum_\beta \Bigr [\bar z_\alpha (\bar r_\beta-r_\beta )-\bar r_\alpha (\bar z_\beta-
z_\beta) \Bigr] 
\nonumber\\
&\times [1-\cos \theta_{\alpha ,\beta }]
-(\eta \bar z_\alpha  - \xi \bar r_\alpha) /|{\bf p_\alpha}|+G_F n_e \bar r_\alpha \,,
\nonumber\\
\,
\label{eom1}
\end{eqnarray}
and for the flavor conserving operators by,
\begin{eqnarray}
&i {d \over dt} z_\alpha=-{ \sqrt{2} G_F n_\nu\over N_{B}}\sum_\beta  \Bigr[ r_\alpha^*(\bar r_\beta -r_\beta) - r_\alpha (\bar r_\beta^*-r_\beta^*)\Bigr]
\nonumber\\
&[1-\cos \theta_{\alpha ,\beta }]  
+2 \eta (r_\alpha -r^*_\alpha)/|{\bf p_\alpha}|\, ,
\nonumber\\
\,
\nonumber\\
&i {d \over dt} \bar z_\alpha={ \sqrt{2} G_F n_\nu\over N_{B}}\sum_\beta  \Bigr[ \bar r _\alpha^*(r_\beta -\bar r_\beta) -\bar r_\alpha (r_\beta^*-\bar r_\beta^*) \Bigr] 
\nonumber\\
&[1-\cos \theta_{\alpha ,\beta }]
- 2 \eta (\bar r_\alpha -\bar r^*_\alpha) /|{\bf p_\alpha}| \, .
\nonumber\\
\,
\label{eom2}
\end{eqnarray}
Note that in the equations (\ref{eom1}), (\ref{eom2}), once we take the mean-field limit, the Hermitian adjoints in the original operator expressions turn into simple complex conjugates of the functions.

We shall use (\ref{eom1}), (\ref{eom2}) to determine the evolution of the system starting from a flavor-diagonal initial state at $t=0$, that is, from $r_\alpha(0)=0$, $\bar r_\alpha(0)=0$. Consider first the system with the parameter $\eta=0$, and the initial values $r_\alpha=0$,
$\bar r_\alpha=0$. Then the solution has values of $z_\alpha$, $\bar z_\alpha$,
that are constant in time, with vanishing $r_\alpha$, $\bar r_\alpha$. We examine the stability
by linearizing around this solution. We note that the linear perturbations of the $z$'s
do not enter (\ref{eom1}) at all, so that we have linear equations for perturbations
$\delta r_\alpha$, $\bar \delta r_\alpha$. Grouping these perturbations into a single
column vector $\delta \Omega_{\alpha,i}$ where $i=1$ picks out $\delta r_\alpha$, and $i=2$
picks out $\bar \delta r_\alpha$, we write the equations in matrix form
as,
\begin{eqnarray}
i {d \over dt} \,\delta \Omega=M\, \delta \Omega \,,
\label{pert}
\end{eqnarray}
where $M$ is a $2 N_{\rm B}$ dimensional square matrix with elements $M_{\alpha,i;\beta,j}$ that we read out from
(\ref{eom1}) as, 
\begin{eqnarray}
M_{\alpha,1;\beta,1}=\sqrt{2} G_F n_\nu N_{B}^{-1}\, \Bigr [ z_\alpha [1-\cos 
\theta _{\alpha,\beta}]
\nonumber\\
-\delta_{\alpha,\beta}\sum _\gamma (z_\gamma-\bar z_\gamma )[1-\cos 
\theta _{\alpha,\gamma}]\Bigr ]- \xi |\bf p_\alpha|^{-1}\delta_{\alpha,\beta} ~ ,
\nonumber\\
\,
\nonumber\\
M_{\alpha,1;\beta,2}=- \sqrt{2} G_F n_\nu N_{B}^{-1}\,z_\alpha [1-\cos 
\theta _{\alpha,\beta}]\,,
\nonumber\\
\,
\nonumber\\
M_{\alpha,2;\beta,1}= \sqrt{2} G_F n_\nu N_{B}^{-1}\,\bar z_\alpha [1-\cos \theta _{\alpha,\beta}] \,,
\nonumber\\
\,
\nonumber\\
M_{\alpha,1;\beta,1}=- \sqrt{2} G_F n_\nu N_{B}^{-1}\, \Bigr [-\bar z_\alpha [1-\cos 
\theta _{\alpha,\beta}]
\nonumber\\
-\delta_{\alpha,\beta}\sum _\gamma (z_\gamma-\bar z_\gamma )[1-\cos 
\theta _{\alpha,\gamma}]\Bigr ]+ \xi |\bf p_\alpha|^{-1} \,\delta_{\alpha,\beta}~.
\nonumber\\
\,
\label{matrix}
\end{eqnarray}
The flavor-diagonal initial conditions determine the $2 \,N_{\rm B}$ parameters $z_\alpha$,
$\bar z_\alpha$ in (\ref{matrix}). The instabilities that are the
focus of the present paper all arise from a complex eigenvalue, $\lambda_i$ of $M$. There are two
categories of instabilities that can arise: 

a) Cases in which there is a complex eigenvalue even
when $\xi=0$. The growth rate of the mode can be some appreciable fraction
of $G_F n_e$, the ``very fast" rate mentioned in the introduction. This
situation only arises when the angular distributions in the initial state are somewhat complex .

b) Cases in which the eigenvalues are real when $\xi=0$ but for the case of non-vanishing $\xi$ are complex with imaginary
part of order $(G_F \xi n_\nu/|{\bf p}|)^{1/2}$, for small $\xi$. This is the ``fast rate" mentioned in the 
introduction, still much faster than the ordinary collision frequency in our region
of interest. 
\section{3. Six beam example}
We have done a number of simulations with different distributions in angle and 
flavor. Simulations including bins of different energies have been done as
well, but it turns out that including an energy spectrum in the dynamics
makes little difference to the results, although energy enters in
the factor $|{\bf p}|^{-1}$ in the oscillation rate. This is basically
the synchronization phenomenon discussed in refs \cite{ks}. Of course, energy enters
in a bookkeeping way, since the main point will be the equalization of 
the $\nu_x$ spectrum and the $\nu_e$ spectrum through flavor trading.

Since a flavor independent distribution $\rho_{i,j}=\delta_{i,j}$
is completely irrelevent in these equations, we can subtract a flavor independent
$n_{\rm FI}$ part, replacing the multiplying $n_\nu$ in (\ref{eom1}) and (\ref{eom2}) by $n_{\rm eff}=(n_{\nu}-n_{\rm FI})$, 
and giving a residual distribution that is all (or nearly all) $\nu_x$'s
going up and  $\nu_e$'s going down.
We begin with one of the simpler cases for which we have carried out numerical
solutions of the evolution equations.
 
The physics is that 
in a critical region the $\nu_x$, $\bar \nu_x$ have an outward bias in their momentum distributions while the 
$\nu_e$, $\bar \nu_e$ are more istropically distributed. Also there is some excess of $\nu_e$ overall , 
coming from the tail of the deleptonization pulse. This excess dies off rather fast
compared to the time scale of total neutrino energy loss from the star. The $\nu_x$, $\bar \nu_x$ also have 
a harder energy spectrum; as noted above, this latter fact is fairly unimportant as far as the mechanics of
flavor exchanges is concerned, but of course it is central to the conclusions being interesting.

For the simulation, take six beams, $N_{\rm grp}=6$, oriented in the $\pm \hat x \,,\pm \hat y \,,\pm \hat z$ directions
in ordinary space, which we label as the $\pm  1 \, \pm 2 \,, \pm 3$ directions.

We take the following values for the functions in (\ref{eom1}), (\ref{eom2}) at $t=0$ :
\begin{eqnarray}
z_{3}(0)=-1~,~\bar z_{3}(0)=-1~,
\nonumber\\
z_{\pm 1}(0)=g ~,~\bar z_{\pm 1}(0)=0~,
\nonumber\\
z_{\pm 2}(0)=g~,~\bar z_{\pm 2}(0)=0~,
\nonumber\\
z_{-3}(0)=1+2g ~,~\bar z_{-3}(0)=1~.
\label{initials}
\end{eqnarray}
Here, the $\hat 3 $ direction is up (i.e., outward) and the choice 
$z_{3} n_{\rm eff}=\rho_{e,e}^{(3)}-\rho_{x,x}^{(3)}=-n_{\rm eff}$ for the density
function for the upward flux indicates an excess of $\nu_x$'s over $\nu_e$'s. Likewise
the part with unity in the down direction reflects the excess of $\nu_e\,, \bar \nu_e$ in 
the downward direction. A distribution that is independent of
particle flavor is sterile and can be subtracted. We categorize the remainder as:

1) In the up direction, all $\nu_x \,,\bar \nu_x$'s.

2) In the transverse directions, equal numbers of $\nu_x \,,\bar \nu_x, \,\bar \nu_e$,
but with an excess of $\nu_e$, measured by the parameter $g$.

3) In the downward direction, all $\nu_e,\,\bar \nu_e$, with a greater excess
of $\nu_e$.

We introduce a dimensionless parameter R to measure the ratio of the vacuum oscillation rate to
the ``very fast" neutral current rate, 
\begin{eqnarray}
R=(\delta m^2)(G_F n_{(0)} |{\bf p}|)^{-1}\sim 10^{-9}
\label{R}
\end{eqnarray}
where the estimate is for the conditions that we describe above, using the solar neutrino value 
$\delta m^2\approx 10^{-4}$.  In our numerical studies, 
we have looked at the region $10^{-8}<R<10^{-2}$, with fixed, physical $\sin^2 (2\theta_{\rm vac})=.86$.

In our first example we choose $\xi$ to be positive, which corresponds to
the ``inverted hierarchy".
We look at the eigenvalues of the matrix $M$ of (\ref{matrix}) in the case $g=.2$, which 
corresponds to a $\nu_e$ surplus of roughly $(\nu_e-\bar\nu_e)/\bar \nu_e$ of
10\% , obtaining, among the twelve eigenvalues, $\lambda_i$ one complex conjugate pair with imaginary part, 
\begin{eqnarray}
{\rm Im} [\lambda ]=.66[G_F\, R] ^{1/2}\,,
\label{scales}
\end{eqnarray}
where the exponent $1/2$ is essentially exact over a region
in which $R$ changes by over six orders of magnitude.

Next we calculate the evolution of the system by solving the twelve equations (\ref{eom1}), (\ref{eom2})  with the initial conditions given by (\ref{initials}), for a range of parameters
$g$ (representing $\nu_e$ excess),  $\xi$, with $\lambda$ determined by keeping the value
$\sin^2 (2\theta)= $ fixed.  

In the example shown, we took an electron density corresponding to matter with
a density of $5\times 10^{11} {\rm g c^{-3}}$, and electron fraction $Y_e=.4$.
We take a neutrino density $n_\nu$ for each species given by a thermal density at a temperature
of $ 7 \,{\rm MeV}$; and for definiteness $n_{\rm eff}=.2 n_\nu$, and $g=.2$ as above.

Again we have calculated the development
as a function of $R$ over the range $10^{-1}>R>10^{-7}$.  
Results are shown in fig. 1 for the case $R=10^{-5}$.  For our range of $R$
we find that the first peak corresponding to total flavor takeover comes at a 
mixing time $t_{\rm mix}$,
\begin {figure}[ht]
    \begin{center}
       \epsfxsize 2.75in
        \begin{tabular}{rc}
           \vbox{\hbox{
$\displaystyle{ \, { } }$
               \hskip -0.1in \null} 
\vskip 0.2in
} &
            \epsfbox{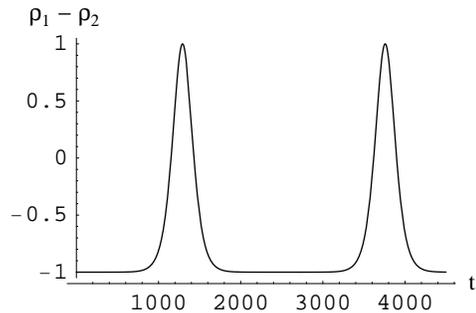} \\
            &
            \hbox{} \\
        \end{tabular}
    \end{center}
\label{fig. 1}
\protect\caption
    {%
Plot of the changing $\nu_e$, $\nu_x$ asymmetry in the upward hemisphere, $\rho_1-\rho_2$,
for the case $R=10^{-4}$ and inverted hierarchy,
where $\langle \rho_{\{1,2\}} \rangle$ are the respective densities of the two
flavors as measured in the units $n_{\rm eff}$. The initial
value, $\rho_1-\rho_2=-1$ indicates total occupancy of our subset of states with $\nu_x$'s.
The unit of time is the inverse of the fast rate $G_F n_{\rm eff}$.
}
\end {figure}
\begin{eqnarray}
t_{\rm mix}^{-1}= G_F n_{\rm eff} [R ]^{.56}\,,
\label{mix}
\end{eqnarray}
where the fit is (within our precision) { \it exact } over six decades as we change $R$.
We have no insight into this law, except insofar as we anticipated something close
to $R^{.5}$ dependence, based on the eigenvalue dependence noted above. Of course we would not
have anticipated an exact $R^{.5}$ law, since the system is completely into a nonlinear
domain by the time of maximal mixing.

Now extrapolating to the value $R=10^{-9}$ appropriate to neutrinos with average number density
corresponding to temperature $T=7~ {\rm MeV}$, and energies of order $20 \,{\rm MeV}$
we find that the first turnover of flavor occurs in a distance $\sim 30$ cm.

Keeping the beam configuration and the initial assignments fixed as above we have varied the parameters
in the solutions to see if there are qualitative sensitivities:

a. Changing the all-over neutrino density, $n_\nu$ ,(i.e. the temperature) or the ratio, $n_{\rm eff}/ n_\nu$,
(i.e. the scale of the initial asymmetry) is unimportant. It is true that with a smaller asymmetry the flavor exchange
will be slower, but for the case of our parameter region the exchange 
will still be fast measured in terms of  other time scales in the problem. In particular the turnover region is 
small compared to the (rather idealized) distance between a $\nu_e$-surface and a $\nu_x$-surface
(of last scatterings).

b. Changing the electron density is also not at all interesting; a change by a factor of  2 typically makes a change of no more than 10\% in the position of the tipping point. 

c. For the case of the ``normal" hierarchy we get instability only for very small $\nu_e$ excess.
This observation agrees completely with the findings of ref. \cite{raf5}.
We would map out the region of instability in more detail, except that the results of the next
section, where the initial angular distributions are taken to be more
complex, provide a much likelier mechanism for rapid mixing in case of the normal hierarchy.

d. Ordinary scattering is very unlikely in the turnover times that correspond to
fig. 1. But we could think of including it to see where the effect of damping
by scattering leaves the distributions after many, many oscillations
of the type shown in fig. 1. Note that in the data shown the time-averaged
occupancies retain most of the initial asymmetry, in spite of the spikes to
reversed occupancy. Taking damping from scattering, following the prescriptions
of McKellar and Thomson \cite{mt} (where, in the monoenergetic idealization the effects
are just a damping term in equations for
the off-diagonal parts of the density matrix), we indeed find near zero asymmetry
and little residual oscillation after only one or two scattering times.
We could not do these calculations using the physical values of the parameters, however,
because this would have required hundreds of the oscillations of which two are shown in
fig. 1. However in the next section we find that with more chaotic initial conditions
we get complete mixing in the time average at shorter times, and without scattering.
\section{4. More complexity}
We move to better coverage, with 14 rays replacing the 6 in the simulations of the preceding section.
To the earlier configuration of rays we add the eight beams that make an angle of $\pi/4$
with the $z$ axis and are positioned in the $x,z$ or $y,z$ planes.
For the four of these beams in the upward hemisphere, we assign 
common initial values $z_u$,$\bar z_u$
that interpolate between the upward and horizontal beams of (\ref{initials}),
 
\begin{eqnarray} 
z_{u}(0)=-.5+\delta/2~,~\bar z_{u}(0)=-1~.
\label{more-ic-a}
\end{eqnarray}
Similarly for the four new beams in the downward hemisphere, with common initial values
$  z_d$, $\bar z_d$, where we interpolate 
between the completely inward ray in (\ref{initials}) and the horizontal
rays to get the common values,
\begin{eqnarray} 
z_{d}(0)=.5+3 \delta/2~,~\bar z_{d}(0)=-.5~.
\label{more-ic-b}
\end{eqnarray}

We easily confirm the
qualitative conclusions of the last system, in this somewhat finer grained
simulation of the same physics.
That is to say, we again find an instability with growth rate proportional to the geometric mean
between the ``very fast" rate $G_F \,n_\nu$ and the ordinary vacuum oscillation 
parameter,  with nearly the same coefficient and limitations of parameter domains, and a detailed
plot recapitulating fig. 1.

Next we put independent random variations in the 28 initial values. Unfortunately,
a whole landscape of new possibilities arise. To quantify this in an example we return to the linearized
instability condition based on the eigenvalues of $M$ of (\ref{matrix}) in our bigger space, but now 
adding independent variations $\Delta z_i(0),  \Delta \bar z_i(0)$,
distributed randomly on the interval  $\{-.05,.05\}$. We begin by turning off the neutrino mass, $\delta m^2=0$. For the case of the electron excess parameter, $\delta=.1$, we find that
somewhat over 50 \% of the time there is at least one 
complex eigenvalue of $M$. Note that the hierarchy 
is not an issue at this point since  $\delta m^2=0$. The $\nu_e$ excess still can matter; when we increase 
the excess $\nu_e$ parameter to $\delta=.2$ the probability of a complex eigenvalue  decreases to around 25\%. 

When we proceed to calculate the evolution from the full equations (\ref{eom1}), (\ref{eom2}),
we find that in almost every case with a complex eigenvalue there is rapid mixing
of flavors among the rays that go upwards and the rays that go downwards, but it is more irregular that the
behavior shown in fig. 1. We show a typical example in fig. 2 for the case $R=10^{-4}$, where results for
both signs of $\delta m^2$ are plotted. 

\begin {figure}[ht]
    \begin{center}
       \epsfxsize 2.75in
        \begin{tabular}{rc}
           \vbox{\hbox{
$\displaystyle{ \, { } }$
               \hskip -0.1in \null} 
} &
           \epsfbox{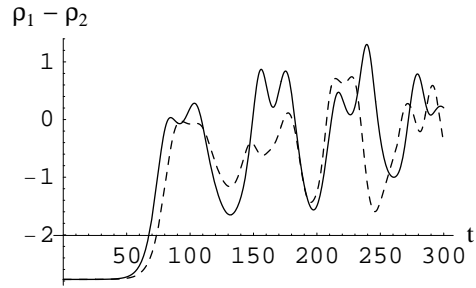} \\
            &
            \hbox{} \\
        \end{tabular}
    \end{center}
\label{fig. 2}
\protect\caption
    {%
For a 14 beam case with the initial conditions as specified in (\ref{initials}),
(\ref{more-ic-a}),(\ref{more-ic-b}), with small (10\%) initial irregularities
introduced so that there is a growing mode in the linearized equations even in the
absence of neutrino mass terms.
Plot of the changing $\nu_e$, $\nu_x$ asymmetry in the upward hemisphere, $\rho_1-\rho_2$
where $\langle \rho_{\{1,2\}} \rangle$ are the respective densities of the two
flavors as measured in the units $n_{\rm eff}$. The initial
value $-1$ indicates total occupancy of our subset of states with $\nu_x$'s.
The unit of time is the inverse of the fast rate $G_F n_{\rm eff}$.
}
\end {figure}

We note that the differences between the normal and inverted
sign are minor. In fig. 3, we show a calculation with the same initial conditions but for the
value $R=10^{-5}$. 

\begin {figure}[ht]
    \begin{center}
       \epsfxsize 2.75in
        \begin{tabular}{rc}
           \vbox{\hbox{
$\displaystyle{ \, { } }$
               \hskip -0.1in \null} 
} &
           \epsfbox{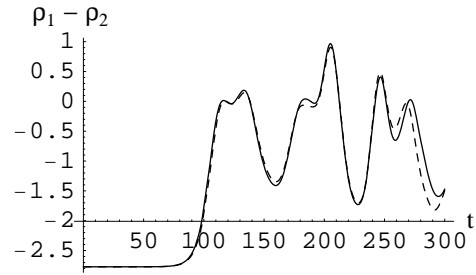} \\
            &
            \hbox{} \\
        \end{tabular}
    \end{center}
\label{fig. 3}
\protect\caption
    {%
The same conditions as shown in fig. 2 but with a neutrino mass parameter
$R=10^{-5}$, instead of $10^{-4}$, showing almost the same behavior for
the normal and inverted hierarchy cases. 
}
\end {figure}
For even smaller values of $R$ all visible 
differences between normal and inverted hierarchies disappear. For the range 
of values $10^{-6}<R<10^{-2}$ we find that (\ref{mix}) can be replaced by,
\begin{eqnarray}
t_{\rm mix}^{-1}= G_F n_{\rm eff} [R ]^{.18}\,,
\label{mix2}
\end{eqnarray}
although the fit is no longer perfect.  In any case the flavor mixing is even faster that that estimated
in the last section, and it is much more thorough. 

As an example we take 
initial conditions defined by beginning with neutrino number densities characteristic of equilibrium
at $T=7$ MeV, then shift 20\% of the down-moving $\nu_x$, $\bar \nu_x$ to
up-moving states, and do the reverse for the $\nu_e$, $\bar \nu_e$ 
distributions. In this case we find total mixing in the time corresponding to $.2$ cm. 
of neutrino path. In general it is clearly not the detailed calculation of evolution,
as shown in fig. 2, that is important, but rather the simple linearized analysis
that detects a growing mode.

The latter depends on the scale of the anisotropies only through 
the fact that the parameter $g$ of (\ref{initials}) measures the
amount of $\nu_e$ relative to the up-down asymmetric part of
the main $\nu,\bar \nu$ distributions. Thus scaled-down anisotropies
in the presence of a big $\nu_e$ excess lead to a smaller chance of
instability, according to our earlier analysis.
\footnote{Of course, the $\nu_e$ excess is very time dependent, decreasing
rapidly during the duration of the neutrino burst.}
We conclude, however, that if the angular distributions are anywhere
nearly as irregular as the usual quantities displayed in the 
results of simulations of the core physics (e.g. $Y_e$), then
there should be many opportunities for instability.
\section{5. Examination of the the mean field approximation.} 
Although we don't find it obvious from the literature, we believe that
getting from the underlying field theory to the evolution equations for the
density matrices involves two separate steps that need to be examined
separately.

1.)Replacing the Hamiltonian by the ``forward" Hamiltonian of (\ref{ham}). 

2.)Making the ``mean field" assumption (\ref{mf}).

The first question is: why do we believe that it is legitimate to ignore almost
all of the terms in the neutrino-neutrino interaction that correspond (in Born approx.)
to scattering at a finite angle? Clearly it is because we are looking for
index of refraction effects, in a sense of order $G_F$, rather than scattering,
where cross-sections are of order $G_F^2$. 

This oversimplifies;
the domains do overlap.
For example, in considering neutrinos passing through an
inhomogenious sea of electrons, density fluctuations on a scale much
greater than the than the $\nu$ wavelength, 
it is correct to scatter from a space varying index of refraction (coming from
the ``forward" interaction Hamiltonian) in place of explicitly adding up waves coming
from the small angle scattering on each electron. But are we sure that when neutrinos
are the target and are themselves subject to flavor manipulation from the beam, the  
index of refraction matrix from forward processes embodies all of the physics? 
Or could small angle (rather than forward) terms enter the analysis
in some nastier way?

That said, and accepting the applicability of the ``forward" Hamiltonian, over time
periods for which scattering is small,
we can ask about the validity of the mean field approximation. 
The essence of the question is captured using a
truncated form of
the neutral current interaction. We return to the effective Hamiltonian (\ref{ham2}) and
discard the oscillation terms and all antiparticle operators, and also insert a parameter
$\eta$ multiplying the term in which the operators do not change the flavor ,
\begin{eqnarray}
H=G \sum_{\alpha,\beta} \Bigr [ 2 r_\alpha^\dagger r_\beta+\eta z_\alpha z_\beta]
 [1-\cos\theta_{\alpha,\beta}]\,.
\label{ham-trunc}
\end{eqnarray}

We take only an up beam A and a down beam B , and
change to the notation appropriate to a system of N spin 1/2 particles
in group A
and N in group B,
 
\begin{eqnarray}
z_{1}=\sum_{i \, \subset A} \sigma_3^i~~,~~z_{2}=\sum_{i \, \subset B} \sigma_3^i~~,~~
\nonumber\\
\,
\nonumber\\
r_{1}=\sum_{i \, \subset A} \sigma_{(+)}^i~~,~~r_{2}=\sum_{i \, \subset B} \sigma_{(+)}^i~.
\end{eqnarray} 
The (mean field) evolution equations are,
\begin{eqnarray}
i {d \over dt} r_1=G(z_1 r_2 -\eta r_1 z_2 )
\nonumber\\
i {d \over dt} r_2=G(z_2 r_1 -\eta r_2 z_1 )\, ,
\label{demo1}
\end{eqnarray}
and
\begin{eqnarray}
i {d \over dt} z_1=G(r_1 r^*_2 - r_1^* r_2 )
\nonumber\\
i {d \over dt} z_2=G(r_2 r^*_1 - r^*_2 r_1 )\, .
\label{demo2}
\end{eqnarray}

We take initial $z_1(0)=N$, $r_1(0)=0$(spins up) ; 
$z_2(0)=-N  $; $r_2(0)=0 $ (spins  down).
The eigenvalues that determine stability in the linearized system are easily read off 
from (\ref{demo1}), $\lambda=\pm G N  \sqrt{\eta^2-1}$.
First we consider the case $\eta=0$ where the eigenvalue shows instability.
We have two groups of spins with no intragroup
interactions but for each pair, one from one group and one from the other, an interaction
of the ``x-y" model form \cite{rfs3}.
Now we take the initial values of the spins in group \#1 all to point in a direction a very
small angle $\alpha$ from the the $\hat z$ axis, and the spins from group \#2 to point 
exactly down. The question is: what is the time scale for spin mixing from the above
mean field equations? Since the growth of the off-diagonal operator
is as,
\begin{eqnarray}
r_1\sim e^{GNt}\, \alpha
\end{eqnarray}
The answer is $t_{\rm mix}\approx (GN)^{-1} ({\rm log }\, \alpha)^{-1}$.

Next we ask what happens when $\alpha=0$, where the mean field
equations say there is no evolution at all. In ref.\cite{rfs3}
we reported complete calculations based on (\ref{ham-trunc}) for cases of N=512 (and fewer) spins in each group.
We found 
excellent fits to the form,
\begin{eqnarray}
t_{\rm ev}={\rm evolution~ time ~ to ~50\% ~mixed}=  (GN)^{-1} {\rm log} N
\label{speed-up}\,,
\end{eqnarray}
a result that is backed up by unrigorous analytical work \cite{rfs4}.

Thus although it is technically correct to say that this correction to the mean
field approximation vanishes in the limit of a large number of particles,
it is only by a logarithm, and logarithms are never huge. That said,
the comparison of mean-field with an initial tilt $\alpha$ with the
full $\alpha=0$ solution shows that the initial-tilt produces
more rapid transformations as long as  $\alpha>N^{-1}$.

This is
effectively the case in our present application. \footnote{We assume that
the effective value of $N$ would be the number
of neutrinos in a volume of dimension of some neutrino coherence length,
not that we know exactly how the latter should be defined.}
Therefore in our present application we assume that the mean field assumptions are justified.
We offer a caveat here, however; we have not carried out the real $2N$ spin
solution when oscillations are present, where, for all we know, the outcome
could be different than the mean field case. It is not guaranteed that 
the solution will interpolate in a simple way between the two limiting cases that we
have analyzed.

We leave this problem for the future, but we can sharpen it a little
here. Addressing the model of (\ref{ham-trunc}) in ref. \cite{rfs3},
we also reported results for the case $\eta=1$, where the eigenvalues $\lambda$
are real. Here we found was no speed-up of the form (\ref{speed-up}).
Instead we found, $t_{\rm ev}\sim G^{-1} N^{-1/2}$, slower by a factor of
$\sqrt N/{\rm log}N$. This can be characterized as the perturbative
time, since for small times, the transition
probability for a particular spin is
\begin{equation}
{\rm prob}\sim G^2 N t^2
\end{equation}
Thus our conjecture is that when the linear response calculation gives
an exponentially growing mode, then the real solution gives
speed-up as defined in (\ref{speed-up}). In contrast, when the
eigenvalues, $\lambda$, are real, we get the perturbative rate. 

More evidence for the latter is found in Friedland and Lunardini's 
analytic solution \cite{fl} of a model of the latter category
(real eigenvalues for the perturbations in the mean field equations).
Their model  was basically that of (\ref{ham-trunc}) with $\eta=1$
together with the assumption of isotropy, $[1-\cos\theta_{\alpha,\beta}]\rightarrow 1$.
\footnote{ The same solution technique can be used to solve the
case of  $\eta=1$  with two opposed beams that we 
discussed above, and also the complementary case of $\eta=0$ and
isotropy, in which case the mean field eigenvalues are  
$\lambda=\{0\, ,\, z_1(0)+z_2(0)\}$, real for any  
initial configuration. The present remarks apply to the solutions
of similar models in refs \cite{mck} as well.}
\section{6. Discussion}
We have made a case for rapid spectrum (and angle) mixing for systems that have two kinds of neutrinos
and antineutrinos. We fully expect the case with all three flavors, and more mixing parameters,
to mix rapidly in every direction; a primitive example was worked out in ref. \cite{rfs1}.

To compare with other authors' results in related calculations, first we emphasize
that most published examples of models with flavor exchange depend
on MSW transitions over a distance in which the electron density is changing
significantly. By contrast, our instabilities and rapid flavor exchange do not
depend on changing the electron density and in any event can occur in regions with
parameters that are far from resonance parameters, as in the region just under the
neutrino-surface that we focused on here. The transition distances that we find
in our calculations, of order 1 cm., are also much shorter than the resonance
regions that appear at smaller electron densities.

The basic mechanism at work in our calculation is similar to that found
in refs. \cite{ful2}-\cite{ful5} , which consider rapid flavor exchanges just downstream from
the neutrino-surface. Though it is not so far from the region that we
considered, the neutrino angular distributions are very different in the
two regions. Outside of the neutrino surface all of the flow is outwards
and the relative angles between neutrino trajectories are diminishing
rapidly as one moves outwards. Since these angles are the key to all
$\nu-\nu$ effects (as they are in our region as well) the results
cannot be compared in detail with ours. 

But we can compare general features of the results. The time scale
for the rapid flavor exchanges found in refs. \cite{raf5} and \cite{raf3} is 
the ``fast" rate as defined in our introduction, the geometric
mean of the neutral current rate and the vacuum oscillation rate.
As in our case the phenomenon does not depend on the space variation
of the electron density. Except for the geometry, the physics
appears to be the same as in our ``six beam" example. We confirm
the conclusions of ref. \cite{raf5} as to the critical roles (and linking)
of the $\nu_e$ excess and the normal vs. inverted hierarchy choice.
In the case of normal hierarchy, the exchange occurs only for
very small values of the electron excess. The case of inverted
hierarchy allows considerable electron excess however.

Our results in sec.4, where we considered a less symmetrical
angular distribution, are quite different. When the linearized
stability analysis in the absence of neutrino mass terms shows
a growing mode, then the solution for the complete problem
shows large mixing for both the normal and inverted hierarchy
cases, in the presence of moderate $\nu_e$ excess. The evolution rate
is also an order of magnitude or more than in the more symmetric``six beam"
example.

Although the authors of ref. \cite{raf3} did not find such behavior in the
region that they treated, it is possible that if perfect cylindrical 
symmetry in the initial distributions were given up, then  these super-fast
turnovers would be produced for both the normal and inverted hierarchy.

We could ask how much it matters whether or not the emerging 
energy spectra are partially or fully
homogenized. There is strong evidence in the literature \cite{R1}-\cite{R3} that it would matter significantly
for the R process nucleosynthesis yields, though we believe that there is no single graph showing
the effects with precision, as there is other complex physics involved.
An interesting question is that of whether or not it matters to the actual
explosion dynamics. The neutrino heating in the region above the
neutrino-surfaces is augmented appreciably in a scenario in which $\nu_e$'s get boosted
in average energy by trading spectra with $\nu_x$'s. 

Of course, feeding our detailed considerations into a big code for the 
supernova is out of the question. The hydrodynamics plus
neutrino transport problem has not been formulated in a way that admits
collective effects of the kind we have discussed here, so far as we know;
and there seems little possibility that it could be in a way that allowed
whole-star computations in any reasonable reasonable amount of computer
time.

But it would be easy for the numerical simulators to check the limiting
case of our suggestion, total instantaneous flavor homogenization everywhere
within the outer neutrino-surface, putting it by hand into the codes. 
  
The third consequence, and most obvious observable consequence of our
considerations, is the effect on the neutrino pulse signals from
a nearby supernova. Much has been written on this subject, but the
chance of observational data within the near future is very small,
and loose ends can be cleared up after the fact, since the design of
observing apparatus appears not to depend on the details of the predictions. 

This work was supported 
in part by NSF grant PHY-0455918.

\end{document}